\documentclass[nofootinbib, notitlepage]{revtex4-1}
\pdfoutput=1 
\usepackage{epsfig,amssymb,amsmath,latexsym}

\usepackage{pifont}
\usepackage{graphicx}
\usepackage{dcolumn}
\usepackage{bm}
\usepackage{epsfig}
\usepackage{color}

\usepackage{soul}

\newcommand{\Ha}{{\cal H}}
\newcommand{\bx}{\bm{x}}
\newcommand{\fnl}{{f}_\text{NL}}
\def\k{\bm{k}}
\newcommand{\p}{\partial}
\newcommand{\Q}{\mathcal{Q}}

\newcommand{\rd }{}

\begin{document}

\title{The observed galaxy bispectrum  {from single-field inflation} in the squeezed limit}

\author{Kazuya Koyama$^1$, Obinna Umeh$^{1,2}$, Roy Maartens$^{1,2}$, Daniele Bertacca$^{3,4,5,6}$}

\affiliation{$^1$Institute of Cosmology \& Gravitation, University of Portsmouth, Portsmouth PO1 3FX, UK}
\affiliation{$^2$Department of Physics \& Astronomy, University of the Western Cape, Cape Town 7535, South Africa}
\affiliation{$^3$Dipartimento di Fisica e Astronomia ``G. Galilei'', Universit\`a degli Studi di Padova, via Marzolo 8,  I-35131 Padova, Italy}
\affiliation{$^4$ INFN, Sezione di Padova, via Marzolo 8, I-35131 Padova, Italy}
\affiliation{$^5$ Argelander-Institut f\"ur Astronomie, Auf dem Hugel 71, D-53121 Bonn, Germany}
\affiliation{$^6$ Departament de F\'isica Qu\`antica i Astrofis\'ica \& Institut de Ci\`encies del Cosmos, (ICCUB),  Universitat de Barcelona, Mart\'i i Franqu\`es 1,  08028 Barcelona, Spain}
\date{\today}

\begin{abstract}
Using the consistency relation in Fourier space,
we derive the observed galaxy bispectrum from single-field inflation in the squeezed limit, in which one of the three modes has a wavelength much longer than the other two. This provides a non-trivial check of the full computation of the bispectrum based on second-order cosmological perturbation theory in this limit. We show that gauge modes need to be carefully removed in the second-order cosmological perturbations in order to calculate the observed galaxy bispectrum in the squeezed limit. We then give an estimate of the effective non-Gaussianity due to \rd{general-relativistic} lightcone effects that could mimic a primordial non-Gaussian signal.
\end{abstract}


\maketitle

\section{Introduction}
Future galaxy surveys such as Euclid, LSST and SKA \cite{Abate:2012za,Amendola:2016saw,Maartens:2015mra},  will cover larger and larger scales where general relativistic effects become important. Galaxy counts are distorted due to the fact that we observe them on the past lightcone \cite{Yoo:2009au,Yoo:2010ni,Bonvin:2011bg,Challinor:2011bk,Jeong:2011as,Bertacca:2012tp}. These lightcone effects contain various relativistic corrections, which provide new opportunities to test general relativity on ultra-large scales \rd{(i.e., super-equality scales, where the transfer function is $\sim 1$)} \cite{Hall:2012wd,Lombriser:2013aj, Camera:2013kpa, Raccanelli:2013dza,Bacon:2014uja,Raccanelli:2015vla,Alonso:2015uua,Alonso:2015sfa,Fonseca:2015laa,Bonvin:2015kuc,Gaztanaga:2015jrs,Raccanelli:2016avd,Bonvin:2016dze,Borzyszkowski:2017ayl,Lepori:2017twd}. The observed galaxy power spectrum and bispectrum on ultra-large scales are also sensitive to primordial non-Gaussianity and the failure to include relativistic corrections can bias the measurement of primordial non-Gaussianity \cite{Bruni:2011ta, Creminelli:2011sq,Yoo:2011zc,Bruni:2014xma,
Camera:2014sba,Umeh:2016nuh,DiDio:2016gpd,Abramo:2017xnp,Camera:2018jys}. 

The computation of the tree-level bispectrum requires going up to second order in perturbation theory and including all lightcone effects at second order. These effects have been computed in \cite{Bertacca:2014dra,Bertacca:2014wga,Yoo:2014sfa,DiDio:2014lka,Bertacca:2014hwa}. Recently, the galaxy bispectrum with all local relativistic projection effects, i.e., neglecting terms involving lensing and other line-of-sight integrals, has been computed \cite{Umeh:2016nuh,Jolicoeur:2017nyt,Jolicoeur:2017eyi} (for the formalism in the general case, see \cite{Bertacca:2017dzm}). In Fourier space, these effects start to dominate on super-equality scales for any configuration of the bispectrum \cite{Umeh:2016nuh,Jolicoeur:2017nyt,Jolicoeur:2017eyi}. 

In the squeezed limit, where one of the three modes has a wavelength much longer than the other two, a simple analytical result can be obtained by using the consistency relation \cite{Maldacena:2002vr,Bartolo:2001rt,Valageas:2013cma, Senatore:2012wy, Weinberg:2003sw, Pajer:2013ana,Creminelli:2013nua,
Creminelli:2013poa,Creminelli:2013mca,Kehagias:2013paa,Kehagias:2013rpa,Kehagias:2013yd,Kehagias:2015tda}. The effect of long-wavelength modes on small-scale modes can be regarded as a coordinate transformation in a patch where short-wavelength modes live. This may introduce a correlation between long and short modes, leading to a nonzero squeezed-limit bispectrum. We use this technique to derive the observed fluctuation of the number counts and confirm the validity of the full second-order results in the squeezed limit. 

\rd{This technique for  computing the squeezed-limit bispectrum comes with a warning in the case of Gaussian initial conditions, e.g. in a single-field inflation model. If we use the global coordinates corresponding to accessing the entire volume of the Universe, then we inadvertently include gauge modes (or gauge `artifacts') in the consistency relation. 
These modes, which are not excluded in perturbation theory, have arbitrarily long wavelengths, greater than the maximum observable scale, and should not contribute to observables.
In a single-field inflation model,
these gauge modes can be removed by using local coordinates, which correspond to accessing only a finite volume of the Universe. In this way, we find the correct result for the bispectrum --
 the removal of the gauge modes leads to a {\em vanishing} squeezed-limit bispectrum \cite{Gerstenlauer:2011ti,Tanaka:2011aj,Pajer:2013ana,LoVerde:2013dgp, dePutter:2015vga, Dai:2015jaa, Bartolo:2015qva,Cabass:2016cgp,Bravo:2017gct}, {\em provided} that we neglect the projection effects of observing on the past lightcone. When the lightcone effects are taken into account, as we do here, then the squeezed-limit bispectrum is no longer zero, since the ultra-large scale relativistic effects correlate separated patches. However, it is still
 crucial to remove the gauge modes in second-order perturbations, in order to satisfy the consistency relation in observational coordinates}. 

Using our expression for the observed galaxy number density, which matches the second-order perturbation theory prediction after the gauge modes have been removed, we derive the galaxy bispectrum in the squeezed limit for Gaussian initial conditions.  We compare this to the Newtonian limit of the galaxy bispectrum in the same shape configuration but for non-Gaussian initial conditions in order to estimate the effective local non-Gaussianity due to general relativistic light cone effects. The result we obtain is similar to that of \cite{Kehagias:2015tda} (we discuss the difference between their result and ours in Appendix A).

This paper is organised as follows. In section \ref{sec:numbercount}, we discuss the derivation of the observed galaxy number density and derive the consistency relation. In section \ref{sec:consistencysecondorder}, we compare the consistency relation with the full second-order result in the squeezed limit. In section \ref{sec:spatialdilatation}, we identify the gauge mode that needs to be removed from the second-order cosmological perturbations and show that after removing the gauge mode, the full second-order result satisfies the consistency relation in the squeezed limit. We derive the expression for the observed galaxy bispectrum in the squeezed limit and compute the effective local non-Gaussianity due to general relativistic light cone effects in section \ref{sec:squzzedbispectrum}.  Section \ref{sec:conc} is devoted to conclusions. 

Throughout the paper, we assume that our universe is described by the $\Lambda$+Cold Dark Matter ($\Lambda$CDM) model. We fix the cosmological parameters using the Cosmic Microwave Background (CMB) constraint from the latest Planck experiment \cite{Ade:2015xua}:
$H_0=67.8\,{\rm km/ s/ Mpc},$ $\Omega_{m0} =1-\Omega_{\Lambda 0}= 0.308$. 

\section{Consistency relation for the observed galaxy number density}\label{sec:numbercount}

In this section we derive the observed galaxy number density fluctuations following \cite{Jeong:2011as} (see also \cite{Bonvin:2011bg,Challinor:2011bk}). 
In the perturbed universe, the observed redshift of galaxies  $\tilde{z}$ is given by 
\begin{equation}
1 + \tilde{z} = (1 + \bar{z}) (1 + \delta z),
\end{equation}
where $\bar{z}$ is the redshift in the unperturbed background universe. The observed position of the galaxies $\tilde{\bx}$ that is inferred assuming unperturbed geodesics, is related to the true position of the galaxies $\bx$ through 
\begin{equation}
\bx = \tilde{\bx} + \Delta \bx.
\end{equation}
The intrinsic comoving galaxy number density is defined as 
\begin{equation}
a^3(\bar{z}) n_g (\bx, \bar{z}) = a^3(\bar{z}) \bar{n}_g(\bar{z}) 
\big[ 1 + \delta_g(\bx, \bar{z})  \big],
\end{equation}
where $n_g(\bx, \bar{z})$ is the physical number density of galaxies as a function of the true position of galaxies $\bx$. Expanding $\bar{z}$ to first order in $\delta z$, we obtain
\begin{equation}
a^3(\bar{z}) n_g (\bx, \bar{z}) = a^3(\tilde{z}) \bar{n}_g(\tilde{z}) 
\big[1 + b_e{(\tilde z)} \delta z  \big] \big[ 1 + \delta_g(\bx, \bar{z})  \big], \qquad 
b_e \equiv \frac{ {\partial} \ln (a^3 \bar{n}_g)}{ {\partial} \ln a}\,,
\label{n1}
\end{equation}
where $b_e$ is the galaxy evolution bias.
The observed galaxy density $\tilde{n}_g(\tilde{\bx},\tilde{z})$ is related to the physical number density of galaxies $n_g(\bx, \bar{z})$ to first order as  
\begin{equation}
a^3(\tilde{z}) \tilde{n}_g(\tilde{\bx}, \tilde{z}) = a^3(\bar{z}) 
n_g(\bx, \bar{z}) \big[ 1+  \delta V -3 \delta z - Q \delta L \big]_{(\tilde{\bx}, \tilde{z})}, 
\qquad 
Q{(\tilde z)} \equiv - \frac{ \partial \ln \bar{n}_g{(\tilde z,>L)}}{\partial \ln L},
\label{n2}
\end{equation}
where we assumed a magnitude-limited sample with cumulative luminosity function {$\bar{n}_g{(\tilde z,>L)}$} and magnification bias $Q$, $\delta L$ is the luminosity perturbation and $\delta V$ is the volume perturbation \cite{Bonvin:2011bg}. 
Combining equations (\ref{n1}) and (\ref{n2}), and defining the observed galaxy number density contrast as 
\begin{equation}
\tilde{n}_g(\tilde{\bx}, \tilde{z}) = \bar{n}_g(\tilde{z})  \big[ 1 + \Delta_g(\tilde{\bx}, \tilde{z}) \big],
\end{equation}
we obtain 
\begin{equation}
1 + \Delta_g(\tilde{\bx}, \tilde{z}) = \big[ 1 +\delta_g(\bx, \bar{z}) \big] \big[1 +  (b_e -3) \delta z - Q\, \delta L + \delta V \big]_{(\tilde{\bx}, \tilde{z})}.
\label{firstorder}
\end{equation}

Now we consider the effect of long-wavelength perturbations on the small-scale observed galaxy number density. The effect of long-mode perturbations is to introduce fluctuations of observed coordinates of the small-scale perturbations, 
\begin{equation}
1 + \tilde{z}= (1+ \bar{z})(1 + \delta z_L), \quad 
\bx = \tilde{\bx} + \Delta \bx_L\,,
\label{long}
\end{equation}
as well as the luminosity and volume perturbations. In the presence of long-wavelength modes, the observed galaxy density contrast on small scales  {$\big[\Delta_{gS}\big]{}_L$} is given by  
\begin{equation}
1 +\big[\Delta_{gS}\big]{}_L(\tilde{\bx}, \tilde{z}) = \big[1 + \Delta_{gS}(\bx,\bar{z}) \big]\big[1 + (b_e -3) \delta z_L - Q\, \delta L_L +\delta V_L \big]_{(\tilde{\bx},\tilde{z})}.
\label{longshort}
\end{equation}
Note that $\Delta_{gS}$ on the right hand side is evaluated at $(\bx, \bar{z})$, which introduces the effect of long modes when they are converted to the observed coordinates $(\tilde{\bx}, \tilde{z})$ through \eqref{long}.

By expanding $\big[\Delta_{gS}\big]{}_L$ to first order in the long-wavelength perturbation, we obtain the second-order observed small-scale galaxy number density in the presence of long modes, in terms of the observed coordinates $(\tilde{\bx}, \tilde{z})$:
\begin{equation}
\big[\Delta^{(2)}_{gS}\big]{}_L= \Delta_{gS} \big[ (b_e-3) \delta z_L - Q\,  \delta L_L + \delta V_L \big]  - (1 + z) \delta z_L \frac{\partial}{\partial z} \Delta_{gS} + \Delta x_{L\|} \partial_{\|} \Delta_{gS} +  {\Delta x_{L\perp}^i \partial_{\perp i}} \Delta_{gS} +  \delta L_L  \frac{\partial }{ \partial L}\Delta_{gS} \,,
\label{consistency}
\end{equation}
where here and below {\em we omit the tilde from observed coordinates}  to simplify the expressions.
We decomposed $ \Delta x^i$ into {components} parallel and perpendicular to the line of sight, $\Delta x^i = n^i \Delta x_{\|} 
+\Delta x_{\perp}^i$, and decomposed the spatial derivative accordingly :  

\begin{equation}
\partial_{\|} = n^i \frac{\partial}{\partial x^i}, \qquad  {\partial_{\perp i}} = \frac{\partial}{\partial x^i} -  {n_i} \partial_{\|},
\end{equation}
where $n^i$ is the unit vector parallel to the line of sight. 

The luminosity and volume perturbations are gauge invariant and thus can be evaluated in any gauge. Here we give the expressions in the Poisson gauge, where the line element is given {in the vanishing anisotropic stress tensor limit} by 
\begin{equation}
ds^2 = a^2(\eta) \Big[- (1 +2 \Phi) d \eta^2 +(1 - 2 \Phi)\delta_{ij} dx^i dx^j \Big]. 
\end{equation}
{The fluctuations of the luminosity and the volume are given by {\cite{Bertacca:2014hwa}}
	\begin{align}
	\delta L&= - 2 \Phi + 2 \Big(1 - \frac{1}{\Ha \chi}  \Big) 
	(-\Phi + \partial_{\|} v + 2 I )- \frac{2}{\chi} T - 2 \kappa,
	\label{luminosity} \\
	\delta V  &= - 4 \Phi + \frac{1}{\Ha} \Phi' +\Big(\frac{\Ha'}{\Ha} + \frac{2}{\chi \Ha}  \Big) \Phi 
	+ \Big(-3 + \frac{\Ha'}{\Ha} + \frac{2}{\chi \Ha}   \Big)(- \partial_{\|}v - 2 I)
	- \frac{2}{\chi} T - 2 \kappa,
	\label{volume}
	\end{align}
	where 
	\begin{align}
	T &= - 2 \int_0^\chi d \hat{\chi} \Phi,  \\
	I & = - \int_0^{\chi} d \hat{\chi} \Phi',\\
	\kappa & =  \int_0^\chi d \hat{\chi} \Big[ (\chi - \hat{\chi}) \frac{\hat{\chi}}{\chi} 
	\nabla_{\perp}^2 \Phi \Big],
	\end{align}
	are the time delay, Integrated Sachs Wolfe (ISW) and convergence, respectively, the prime indicates  partial derivative with respect to $\eta$, and $\chi$ is given by $x^i = \chi n^i$, where $x^i$ are the observed coordinates. Note that we omitted observer-dependent terms. 
	
	On the other hand $\delta z$ and $\Delta \bx$ are gauge dependent, so that we need to specify a gauge. The local galaxy bias should be defined in the rest-frame of cold dark matter, which is assumed to coincide with the rest frame of galaxies on large scales. 
Thus the galaxy bias is naturally formulated in the comoving synchronous (CS) gauge, where  
\rd{the coordinate time is the proper time along matter world-lines and the four-velocity of matter is everywhere orthogonal to the $t=\,$constant hypersurfaces}	
\cite{Challinor:2011bk, Bruni:2011ta, Jeong:2011as, Bertacca:2015mca}. Therefore, we use the CS gauge to obtain the expressions of these quantities: 
	\begin{align}
	\delta z &= -\Phi + \partial_{\|} v + 2 I- \Ha v , 
	\label{redshift}\\
	\Delta x_{\|} & = - \frac{1}{\Ha} (-\Phi + \partial_{\|} v + 2 I)- T  - \partial_{\|} \xi, \\
	{\Delta x_{\perp}^i} & = -2 \int_0^\chi d\chi' \Big[ (\chi - \chi') \frac{\chi'}{\chi} 
	\partial^i_{\perp} \Phi \Big] -  {\partial_{\perp}^i} \xi,
	\end{align}
	where $ {\partial_i}v$ is the Poisson-gauge velocity and $\xi = \int d \eta'  \; v(\eta')$ generates the spatial gauge transformation between the Poisson and CS gauges (see Appendix \ref{sec:appendixA} for details). They 
	give a contribution to the observed galaxy number density of the form  
	\begin{equation}
	- \big(\partial_{\|} \xi_L\big) \big(\partial_{\|} \Delta_{gS}\big)  
	- \big( {\partial_{\perp i}}\, \xi_L\big) \big( {\partial_{\perp}^i} \Delta_{gS}\big) 
	= - \big(\partial_i \xi_L\big)   \big(\partial^i \Delta_{gS}\big).
	\end{equation}
	{However}, in Fourier space, \rd{$\k_S \cdot \k_L  = 0$ to lowest order in $k_L/k_S$}, so that this does not contribute to the consistency relation. 
	
	In the following, we will ignore the integral terms, $T$, $I$ and $\kappa$, as well as the integral term in $\Delta x_{\perp}^i$, in order to compare our results with \cite{Umeh:2016nuh,Jolicoeur:2017nyt, Jolicoeur:2017eyi}. These integral terms can be considered separately. Our expressions for $\Delta x_{\|}$ and $ {\Delta x_{\perp}^i}$ are different from those given in \cite{Kehagias:2015tda} because of an ambiguity in the separation between local and integral terms (see Appendix A for details). 
	
	Using equations \eqref{luminosity}, \eqref{volume} and \eqref{redshift}, \eqref{firstorder} gives the first-order observed galaxy number density contrast:
	\begin{equation}
	\Delta_g= \delta_g  -{\cal H} (b_e-3) v + \Big[
	b_e - \frac{\Ha'}{\Ha} - 2 Q - 2 \frac{(1-Q)}{\chi \Ha}
	\Big] (-\Phi + \partial_{\|} v) + (-1 +2 Q) \Phi
	-\frac{1}{\Ha} \partial_{\|}^2 v + \frac{1}{\Ha} \Phi'.
	\label{firstobs}
	\end{equation}
	We approximate the observed number density on small scales at second order by the Newtonian theory prediction
	\begin{eqnarray}
	\Delta_{gS}^{ (2)}\approx   {\Delta_{g{\rm N}}^{(2)}} &= &\delta_{g}^{(2)}-\frac{1}{\Ha}\partial^2_{\|}v^{(2)} -\frac{1}{\Ha}\delta_{g}  \partial^2_{\|}v
	+\left(\frac{1}{\Ha}\partial^2_{\|}v\right)^2
	-\frac{1}{\Ha}\partial_{\|}v\left[\partial_{\|} \delta_{g}  -\frac{1}{\Ha}\partial_{\|}^3 v \right]\,.
	\label{Newtonian2nd}
	\end{eqnarray}
Up to now, we have omitted the superscript (1) for first-order perturbations, and we will continue to do so except when there is ambiguity. We use the convention $X= X^{(1)} + X^{(2)}$ for second-order perturbations.
	In the Newtonian approximation \eqref{Newtonian2nd}, there is no gauge ambiguity and $\delta_{g}$ may be related to the matter density field via a simple bias relation\footnote{The effect of the tidal bias, \rd{which is expected to be of the same order as that of quadratic bias,} is neglected for simplicity.}: $\delta_g =  b_1 \delta 
	+ b_2\, \delta^2$, where $b_1$ and $b_2$ are the linear and \rd{quadratic galaxy biases in the `local-in-mass-density' bias model}.

\section{The observed galaxy density in comoving-synchronous gauge}\label{sec:consistencysecondorder}

\subsection{Consistency relation}
We decompose $\big[\Delta^{(2)}_{gS}\big]{}_L$ given by \eqref{consistency} into six separate components, following the notation of \cite{Umeh:2016nuh,Jolicoeur:2017nyt, Jolicoeur:2017eyi}:
{\begin{equation}
 \big[\Delta^{(2)}_{gS}\big]{}_L =\big[\Delta_{gS}\big]{}_{L}^{\Gamma_7} +\big[\Delta_{gS}\big]{}_{L}^{\Gamma_8}
+\big[\Delta_{gS}\big]{}_{L}^{\Gamma_9} +\big[\Delta_{gS}\big]{}_{L}^{\Gamma_{11}} +\big[\Delta_{gS}\big]{}_{L}^{\Gamma_{12}} +\big[\Delta_{gS}\big]{}_{L}^{\Gamma_{13}} \,, \label{dgsl}
\end{equation}
where {
\begin{align}\nonumber
\big[\Delta_{gS}\big]{}_{L}^{\Gamma_7} 
&= \Big[-1 -b_e + 4 Q + \frac{2}{\chi \Ha}(1-Q)+ \frac{\Ha'}{\Ha^2}  \Big]
\Phi_{L} \delta_{gS}  + \frac{1}{\Ha}  \delta_{gS}{\Phi_L'} - \frac{1}{\Ha} \Phi_{L} \delta_{gS}'
 - (b_e-3) \Ha v_{L} \delta_{gS} + v_{L} \delta_{gS}'
\\  & \label{eq:Gamma7}
 -2 \Big(2- \frac{1}{\chi \Ha}\Big)\frac{\p   \delta_{gS}}{\p \ln L} \Phi_{L}
\,, \\ \nonumber
\big[\Delta_{gS}\big]{}_{L}^{\Gamma_8} 
&= \frac{1}{\Ha} 
\Big[1+ 2 b_e - 6 Q - \frac{4}{\chi \Ha}(1- Q) - 3 \frac{\Ha'}{\Ha^2}  \Big]
\Phi_{L} \partial_{\|}^2 v_S - \frac{1}{\Ha^2} {\Phi_L'}\partial_{\|}^2 v_S  
+\frac{1}{\Ha^2} \Phi_{L} \partial_{\|}^2 v_S' + (b_e-3) v_{L} \partial_{\|}^2 v_S 
\\  \label{eq:Gamma8} &
 - \frac{\Ha'}{\Ha^2}
v_{L} \partial_{\|}^2 v_S  + \frac{1}{\Ha} v_{L}  \partial_{\|}^2 v_S' \,, \\
\big[\Delta_{gS}\big]{}_{L}^{\Gamma_9} &= \frac{1}{\Ha} \Phi_{L} \partial_{\|} \delta_{gS}\,, \label{eq:Gamma9}\\
\big[\Delta_{gS}\big]{}_{L}^{\Gamma_{11}} 
& = \Big[b_e - 2 Q + 2(Q-1) \frac{1}{\chi \Ha} - \frac{\Ha'}{\Ha^2} \Big]  
\delta_{gS} \partial_{\|} v_{L} + \frac{1}{\Ha} \partial_{\|} v_{L} \delta_{gS}' + 2 \Big(1- \frac{1}{\chi \Ha}\Big) \frac{\p \delta_{gS}}{\p \ln \bar L}\partial_{\|} v_{L}\, \label{eq:Gamma11} \,, \\
\big[\Delta_{gS}\big]{}_{L}^{\Gamma_{12}} &= -\frac{1}{\Ha^2} \Phi_{L} \partial_{\|}^3 v_S \,, \label{eq:Gamma12}\\
\big[\Delta_{gS}\big]{}_{L}^{\Gamma_{13}} &= 
\frac{1}{\Ha} \Big[  
-2 b_e+ 4 Q - 4 (Q-1) \frac{1}{\chi \Ha} + 3 \frac{\Ha'}{\Ha^2} \Big]
\partial_{\|} v_{L} \partial_{\|}^2 v_{S}
- \frac{1}{\Ha^2} \partial_{\|} v_{L} \partial_{\|}^2 v_S'\,. 
\label{eq:Gamma13}
\end{align}}
At second order, we need only the first-order galaxy bias relation to map the galaxy density in \eqref{eq:Gamma7}--\eqref{eq:Gamma13} to the matter density:  {$\delta_{gS} =\delta_{gS}^{(1)} = b_1 \delta_S^{(1)}$.} 

Using the following equations of motion,
\begin{align}
	\Phi' &= - \Ha \Phi + f \Ha \Phi, \quad f= \frac{d \ln \delta}{d \ln a}\,, \\
	\delta_g^{S \,\prime} & = \big(b_1' + b_1 f \Ha \big) \delta \,, \\
	v' &= - \Ha v - \Phi\,,
	\label{Euler}
\end{align}
as well as the background equation,
\begin{align}
	\frac{\Ha'}{\Ha^2} = 1- \frac{3}{2} \Omega_m,  \qquad  \Omega_m = \frac{8 \pi G a^2 \rho_m}{3 \Ha^2}\,,\label{back-relations}
\end{align} 
  we can rewrite \eqref{eq:Gamma7}, \eqref{eq:Gamma8}, \eqref{eq:Gamma11} and \eqref{eq:Gamma13} as follows [\eqref{eq:Gamma9} and \eqref{eq:Gamma12} are unchanged]:
\begin{align}\nonumber
\big[\Delta_{gS}\big]{}_{L}^{\Gamma_{7}}  
&= \Big\{ b_1 \Big[-2-b_e + 4 Q + \frac{2}{\chi \Ha}(1- Q) + \frac{\Ha'}{\Ha^2} \Big] - \frac{b_1'}{\Ha} -2 \Big(2- \frac{1}{\chi \Ha}\Big)\frac{\p   b_1}{\p \ln L}\Big\} \delta_S \Phi_L 
 \\ & -
  \Ha \Big[
b_1(b_e -3 +f) + \frac{b_1'}{\Ha}
\Big ] v_L \delta_S\,, \\
\big[\Delta_{gS}\big]{}_{L}^{\Gamma_{8}} 
& = \frac{1}{\Ha} 
\Big[- 2 f + 2 b_e - 6 Q - \frac{4}{\chi \Ha} (1- Q) - 3 \frac{\Ha'}{\Ha^2}
\Big] \Phi_L \partial_{\|}^2 v_S - \frac{1}{\Ha^2} \Phi_L \partial_{\|}^2 \Phi_S
+ (b_e-3) v_L \partial_{\|}^2 v_S - 2 v_L \partial_{\|}^2 v_S \,,
\label{gamma81}
\\
\big[\Delta_{gS}\big]{}_{L}^{\Gamma_{11}} 
&= \Big\{b_1 \Big[f + b_e- 2 Q + 2 (Q-1) \frac{1}{\chi \Ha}- \frac{\Ha'}{\Ha}  \Big] + \frac{b_1'}{\Ha} +2 \Big(1- \frac{1}{\chi \Ha}\Big)\frac{\p   b_1}{\p \ln L}\Big\} \delta_S \partial_{\|} v_L\,, \\
\big[\Delta_{gS}\big]{}_{L}^{\Gamma_{13}} &= 
\frac{1}{\Ha} \Big[  
1 -2 b_e+ 4 Q - 4 (Q-1) \frac{1}{\chi \Ha} + 3 \frac{\Ha'}{\Ha^2} \Big]
\partial_{\|} v_L \partial_{\|}^2 v_S
	+ \frac{1}{\Ha^2} \partial_{\|} v_L \partial_{\|}^2 \Phi_S\,. 
\end{align}
Here we assumed that the velocity of galaxies is the same as the velocity of dark matter that appears in \eqref{Euler}.

These solutions can be written explicitly in terms of $\Phi$ using the solutions for $\delta$ and $v$ in Fourier space:
\begin{equation}
\delta = - \frac{2}{3} \frac{k^2}{\Omega_m {\cal H}^2} \Phi\,, \quad 
v=   - \frac{2 f}{3 \Omega_m \Ha} \Phi \,. 
\label{first}
\end{equation}
In Fourier space, \rd{all the terms on the right-hand side of} \eqref{dgsl} survive in the limit $k_L\rightarrow 0$. Thus these second-order perturbations contribute to the squeezed-limit bispectrum. See section \ref{sec:squzzedbispectrum} for further details.

\subsection{Second-order perturbations}
The second-order observed galaxy number density was computed in \cite{Bertacca:2014dra,Bertacca:2014wga,Yoo:2014sfa,DiDio:2014lka,Bertacca:2014hwa}.
We use the result presented in \cite{Bertacca:2014hwa} [Eq.~(100)]. The full expression for the number density at second order is composed of two terms: 
\begin{equation}
 {\Delta_g^{(2)} = \Delta_g^{(2,0)}} + \Delta_g^{(1,1)},
\end{equation}
where $ {\Delta_g^{(2,0)}}$ has exactly the same structure as \eqref{firstobs}, with all perturbations being second order, and $\Delta_g^{(1,1)}$ is  {composed of products} of two first-order perturbations. $\Delta_g^{(1,1)}$ gives exactly the same contributions as   {$\big[\Delta_{gS}\big]{}_{L}^{\Gamma_{I}}$} for $I=7,9,11,12,13,$ while for $I=8$ it gives \cite{Jolicoeur:2017nyt}
\begin{align}
	 {\big[\Delta_g^{(1,1)}\big]{}^{\Gamma_8}} 
	= \frac{1}{\Ha} 
	\Big[   
	1 - 2 f + 2 b_e - 6 Q - \frac{4}{\chi \Ha} (1- Q) - 3 \frac{\Ha'}{\Ha^2}
	\Big] \Phi \partial_{\|}^2 v - \frac{1}{\Ha^2} \Phi \partial_{\|}^2 \Phi
	+ (b_e-3) v \partial_{\|}^2 v\,,
	\label{gamma82}
\end{align}
where we neglected the integral terms as we did in our computation of the consistency relation. 
This does not match exactly with \eqref{gamma81}. 
 {The discrepancy between (\ref{gamma81}) and (\ref{gamma82}) should be resolved by the second-order contribution.}

The relevant term in the pure second-order contribution $\Delta_g^{ {(2,0)}}$ in the squeezed limit is 
\begin{equation}
\Delta_g^{ {(2,0)}} = \delta_{g}^{(2) {\rm CS}} - \frac{1}{\Ha} \partial_{\|}^2 v^{(2)},
\label{second}
\end{equation}
where $\delta_{g}^{(2) {\rm CS}}$ is  in the CS gauge while $v^{(2)}$ is in the Poisson gauge. The second term is the second-order Kaiser redshift-space distortion (RSD). 
The full solutions for these perturbations are given by \cite{Villa:2015ppa}, but we focus on the parts that contribute to $\big[\Delta^{(2)}_{gS}\big]{}_L$, which are proportional to $\Phi^2$: 
\begin{align}
\delta^{(2) \rm CS} &= - \frac{4}{3\Omega_m} \left( 1 + \frac{2 f}{3 \Omega_m}   \right) \frac{k^2}{ {\cal H}^2} \Phi^2,
\label{deltaCS}
 \\ 
v^{(2) \rm P} & =  {v^{(2)}} = - \frac{ f}{{\cal H} \Omega_m} \Phi^2. 
\label{vP}
\end{align}
If we assume a  {simple} bias relation  {$\delta_g^{(2)\rm CS} = b_1 \delta^{(2)\rm CS}+b_2\big[\delta^{(1)\rm CS} \big]{}^2$}, then \eqref{second} gives an additional contribution to {$\big[\Delta_{gS}\big]{}_{L}^{\Gamma_{7}}$}, which spoils the agreement with the consistency relation. Also the contribution from (\ref{second}) does not solve the disagreement for {$\big[\Delta_{gS}\big]{}_{L}^{\Gamma_{8}}$}.

In the next section, we derive the solutions for second-order perturbations in the squeezed limit using the consistency relation, and thus identify the gauge mode, which should be removed from (\ref{deltaCS}) and (\ref{vP}) before applying them in \rd{\eqref{gamma82} and} (\ref{second}). 

\section{Second-order perturbations in the squeezed limit}\label{sec:spatialdilatation}
The second-order contribution $\Delta_g^{(2)}$ was computed in~\cite{Bertacca:2014hwa}. \rd{In the squeezed limit, we need to carefully remove the large-scale gauge mode  in  second-order perturbations with Gaussian initial conditions.  
As discussed in \cite{Dai:2015jaa, dePutter:2015vga, Bartolo:2015qva}, this procedure makes  the use of ``short-long" splitting, where local coordinates are defined within a patch whose size is much larger than the typical halo but much smaller than the maximum observable scale.
The curvature perturbation on scales much larger than the patch (but smaller than the observable scale) modulates the patch matter overdensity, leading to a relativistic correction with an effective $f_{\rm NL} ^{\rm GR}=-5/3$. However, this effect cancels out in the halo or galaxy overdensity, when perturbations are evaluated at a fixed local scale, rather than fixed global scale.
Indeed, the small-scale density at a fixed local  physical scale is independent of the long-wavelength perturbation and the long-mode has no effect on the small-scale variance of the halo or galaxy density field smoothed on a fixed mass scale. There is no coupling of short- and long-modes in the absence of primordial non-Gaussianity -- apart from the coupling induced by projection (lightcone) effects.}

In this section, we derive the second-order perturbations in the CS and Poisson gauges using the consistency relation \rd{in {\em real} space, as opposed to Sec. \ref{sec:consistencysecondorder}, where we used the consistency relation in observational space}. In this way we can identify the gauge mode in the second-order perturbations and remove it \rd{in order to resolve the discrepancy found above in \eqref{gamma82} and (\ref{second}).} 

We start from the first-order perturbations. The line element is given by 
\begin{equation}
ds^2 = a(\eta)^2 \Big[   
- (1 + 2 \Psi) d \eta^2 + 2 B_i dx^i d \eta + (1 - 2 \Phi) \delta_{ij} dx^i dx^j 
+ 2 E_{ij} dx^i dx^j \Big]\,, 
\end{equation}
 {where $B_i=\partial_iB$ and $E_{ij} =\big(\partial_i \partial_j - \frac{1}{3} \delta_{ij} \nabla^2 \big) E.$}
Under the gauge transformation
\begin{equation}
\tilde{x}^{\mu} = x^{\mu} - \xi^{\mu}, \quad \xi^{\mu}= (\alpha, \xi^i)\,,
\label{trans0}
\end{equation}
the metric perturbations transform as 
\begin{align}
\tilde{\Psi} &= \Psi + {\cal H} \alpha + \alpha', \\
\tilde{B}_i &= B_i + \xi_i' - \alpha_{,i}\,, \\
-\tilde{\Phi} \delta_{ij} + \tilde{E}_{ij} &= 
-\Phi  \delta_{ij} + E_{ij} + {\cal H} \alpha \delta_{ij} + \frac{1}{2} 
(\xi_{i,j} + \xi_{j,i}), 
\end{align}
while the matter perturbations transform as 
\begin{align}
\tilde{v}_i + \tilde{B}_i &= v_i + B_i - \alpha_{,i}\,, \\
\tilde{\delta} &= \delta  {-} 3 {\cal H} \alpha \,. 
\end{align}
The comoving curvature perturbation is defined as 
\begin{equation}
\zeta \equiv - \Phi + {\cal H} (v - B) = - \left(1 + \frac{2 f}{3 \Omega_m} \right) \Phi,
\label{comcurv}
\end{equation}
where we used the solution (\ref{first}) for $v$ and $B=0$ in the Poisson gauge.  

Now we consider the transformation \cite{Weinberg:2003sw,Tram:2016cpy} 
\begin{align}
\tilde{\eta} &= \eta + \epsilon(\eta), 
\label{trans1}
\\
\tilde{x}^i &= (1+ \lambda) x^i \,,
\label{trans2}
\end{align}
where $\lambda$ is constant. The field equations in the Poisson gauge are invariant under these transformations in the long-wavelength limit. Thus these transformations generate long-wavelength perturbations in the Poisson gauge that automatically satisfy the field equations. To see this fact explicitly, we consider the perturbations generated by these transformations: 
\begin{align}
\Psi  &= - {\cal H} \epsilon- \epsilon'\,, \quad
\Phi = {\cal H} \epsilon + \lambda, \quad \zeta = - \lambda\,, 
\label{phi} \\
\delta &= 3 {\cal H} \epsilon, \quad
 {v   = \epsilon\,, \quad B=0\,.}
\label{delta}
\end{align}
Using equations (\ref{comcurv}) and (\ref{phi}), we find
\begin{equation}
{\cal H} \epsilon = - \frac{2 f}{3 \Omega_m} \Phi. 
\end{equation}
The metric perturbations (\ref{phi}) and matter perturbations (\ref{delta}) indeed agree with the solutions in the Poisson gauge in the long-wavelength limit. The temporal gauge transformation in (\ref{trans1}) corresponds to the transformation from the CS gauge to the Poisson gauge. If we set $\epsilon=0$, these transformations generate long-wavelength perturbations in the CS gauge. 

Consider now the second-order perturbations. First we derive the second-order density perturbations in the CS gauge by setting $\epsilon=0$. The spatial dilatation (\ref{trans2}) generates a density perturbation at second order; \rd{the general form of the transformation is given in  \cite{Malik:2008im}, which leads to}
\begin{equation}
\delta^{ {(2)}(\lambda)} = \lambda \, x^i \frac{\partial}{\partial x^i} \delta   = \left(3 + \frac{ \partial \log \delta}{\partial \log k} \right) \zeta \delta. 
\end{equation}
Using the solution (\ref{first}) for $\delta$ and assuming a scale-invariant power spectrum for $\Phi$, we obtain 
\begin{equation}
\delta^{ {(2)}(\lambda)} = - \frac{4}{3\Omega_m} \left( 1 + \frac{2 f}{3 \Omega_m}   \right) \frac{k^2}{ {\cal H}^2} \Phi^2. 
\end{equation}
This agrees with the squeezed limit of the second-order density in the CS gauge, i.e. (\ref{deltaCS}). 

Next we derive the velocity in the Poisson gauge at second order. The transformations (\ref{trans0}) generate a second-order velocity from the first order velocity; \rd{using \cite{Malik:2008im} again\footnote{\rd{Note that there is a typo in Eq. (6.27) of \cite{Malik:2008im}: the fifth term on the right-hand side should be $+2\alpha_1(v_{1i}'-{\cal H} v_{1i})$.}}}, we find that
\begin{equation}\label{v2g}
 v_i^{ {(2)}} = \alpha (v_i' - {\cal H} v_i) + (v_{i,j} \xi^j - v^j \xi_{i,j}). 
\end{equation}
The first contribution coming from the temporal transformation (\ref{trans1}) gives
\begin{equation}
v^{ {(2)}(\epsilon)} = \frac{ f}{ 3 {\cal H} \Omega_m} \left(\frac{4 f}{3 \Omega_m}  - 1  \right) \Phi^2. 
\end{equation} 
The second contribution on the right-hand side of \eqref{v2g}, from the spatial dilatation (\ref{trans2}), is given by 
\begin{equation}\label{v2ga}
v_i^{ {(2)}(\lambda)} = -  \frac{f}{3{\cal H} \Omega_m} \left(
1 + \frac{2 f}{3 {\Omega_m}} \right){\partial\over\partial x^i}
\left[\Phi^2 - x^j \frac{\partial}{\partial x^j} \big(\Phi^2\big) \right]. 
\end{equation}
Assuming again scale invariance of $\Phi$, we obtain 
\begin{equation}\label{v2gb}
v^{ {(2)}(\lambda)} = - \frac{2f}{3{\cal H} \Omega_m} 
\left( 1 + \frac{2 f}{3 \Omega_m}   \right) \Phi^2.
\end{equation}
Combining \eqref{v2ga} and \eqref{v2gb}, the second-order velocity perturbation is obtained as 
\begin{equation}
v^{ {(2)(\epsilon,\lambda)}} = - \frac{ f}{{\cal H} \Omega_m}\, \Phi^2. 
\end{equation}
This agrees with the squeezed limit of the second-order velocity in the Poisson gauge, i.e. \eqref{vP}. 

We are now in a position to identify the gauge mode in the second-order perturbations. The curvature perturbation contains long modes whose wavelengths are larger than the observed region of the universe. These long modes should be absorbed in background coordinates by a spatial dilatation $\tilde{x}^i = (1 + \zeta) x^i$. This is equivalent to removing the second-order perturbations generated by the spatial dilatation, $\delta^{ {(2)}(\lambda)}$ and $v^{ {(2)}(\lambda)}$. Then, neglecting terms of order $k^2_{L}/k_{S}^2$, the second-order perturbations in the squeezed limit become 
\begin{equation}\label{second-order-SL}
 \delta^{(2) {\rm CS}} =0 , \quad  v^{(2)\rm P}  = v^{ {(2)}(\epsilon)} = \frac{ f}{ 3 {\cal H} \Omega_m}  \left(\frac{4 f}{3 {\Omega_m}}  - 1  \right) \Phi^2. 
\end{equation}
 The second-order RSD gives a contribution to $\big[\Delta_g^{(1,1)}\big]{}^{\Gamma_8}$:
\begin{equation}
 {\big[\Delta_g^{(1,1)}\big]{}^{\Gamma_8}_{\rm RSD}} = - \frac{1}{ {\cal H}}  \partial^2_{\|} v^{(2)} 
= \frac{2 f}{3 {\cal H}^2 \Omega_m} 
\left(1- \frac{4 f}{3 {\Omega_m}} \right) \Phi \partial^2_{\|} \Phi. 
\end{equation}
Adding this contribution to $\big[\Delta_g^{(1,1)}\big]{}^{\Gamma_8}$, we can show that the second-order result agrees with the consistency relation, by using 
\begin{equation}
\left(- \frac{1}{{\cal H}} \Phi  - 2 v \right) \partial_{\|}^2 v 
= \frac{2 f}{3 {\cal H}^2 \Omega_m} 
\left(1- \frac{4 f}{3 \Omega_m} \right) \Phi \partial^2_{\|} \Phi.
 \end{equation}
 
 \section{Observed Galaxy Bispectrum in the squeezed limit}\label{sec:squzzedbispectrum}
In this section, we compute the observed galaxy bispectrum in the squeezed limit in a single-field inflation universe, {following \cite{Umeh:2016nuh}}. 
The bispectrum of the observed galaxy number counts  {at fixed redshift} is defined by
\begin{eqnarray} \label{eq:bispectrumdef}
\big\langle \Delta_{g}( \mathbf{k}_{1}) \Delta_{g}( \mathbf{k}_{2}) \Delta_{g}( \mathbf{k}_{3}) \big\rangle &=& (2\pi)^{3}B_{g}(  \mathbf{k}_{1},  \mathbf{k}_{2},  \mathbf{k}_{3})\delta^{D}( \mathbf{k}_{1}+ \mathbf{k}_{2}+ \mathbf{k}_{3})
\\
&=&\big\langle \Delta_{g}^{(1)}( \mathbf{k}_{1}) \Delta_{g}^{(1)}( \mathbf{k}_{2}) \Delta_{g}^{(2)}(\mathbf{k}_{3}) \big\rangle + \text{2 cyclic permutations} \,.
\end{eqnarray} 
In the squeezed configuration, \rd{to lowest order in $k_L/k_S$,}
\begin{eqnarray} 
&& \k_2 \rd{=} -\k_1\equiv-\k_S,~~\k_3\equiv\k_L,~~ k_1 \rd{=} k_2= k_S\gg k_3= k_L,~~ \k_L\cdot\k_S \rd{=} 0,
\\ &&
 {\mu_S\equiv \mu_1  \rd{=} -\mu_2} , \qquad \mu_L \equiv \mu_3  \rd{=} \sqrt{1- \mu_S^2} \cos \phi \,,
\end{eqnarray} 
where $\phi$ is the azimuthal angle and  $\mu_a=\hat{\bm{k}}_a \cdot {\bm{n}}$. Note that, since we work in Cartesian Fourier space, we use the plane-parallel approximation, and therefore neglect wide-angle effects \cite{Bertacca:2012tp,Jolicoeur:2017nyt}. 

In Fourier space, $\Delta_{g}^{(1)}$ given in \eqref{firstobs} becomes \cite{Umeh:2016nuh} 
\begin{eqnarray}\label{ker1}
\Delta_{g}^{(1)}({\k}) = \mathcal{K}^{(1)}( \k)\delta^{(1)}(\mathbf{k})
\qquad { \rm with} \qquad
\mathcal{K}^{(1)}(\k) =   b_{1} + f \mu^{2} + {\rm i} {\mu\over k} \gamma_1+\frac{\gamma_2}{k^{{2}}},
\end{eqnarray}
where $\gamma_a(z)$ are given in Appendix \ref{sec:appendixB}.
At second order, the kernel for the Newtonian part \eqref{Newtonian2nd} is  \cite{Umeh:2016nuh}
\begin{eqnarray}\label{eq:FourierNewtonian}
{\mathcal{K}^{(2)}_{\rm{N} }(\bm{k}_{1}, \bm{k}_{2},{\k_3})} &=& b_{1}F_{2}(\bm{k}_{1}, \bm{k}_{2}) + b_{2} + fG_{2}(\bm{k}_{1}, \bm{k}_{2})\mu_{3}^{2}
\nonumber\\ &&{}
+  f^2{\mu_1\mu_2 \over k_1k_2}\big( \mu_1k_1+\mu_2k_2\big)^2
+
b_1{f\over k_1k_2}\Big[ \big(\mu_1^2+\mu_2^2 \big)k_1k_2+\mu_1\mu_2\big(k_1^2+k_2^2 \big) \Big],
\end{eqnarray}
where $F_{2}$ and $G_{2}$ are given in Appendix \ref{sec:appendixB}. The Fourier space kernel for $\big[\Delta_{gS}\big]{}_{L} $ follows from the general kernel given in 	\cite{Jolicoeur:2017eyi}:
\begin{eqnarray}\label{eq:kSL}
 {\big[\mathcal{K}_{S}^{(2)}\big]{}_L}(\bm{k}_{1}, \bm{k}_{2}, \bm{k}_{3}) &=& \frac{1}{k_{1}^{2}k_{2}^{2}}\bigg\{ \left(k_{1}^{2} + k_{2}^{2}\right)\Gamma_{7} + \left(\mu_{1}^{2}k_{1}^{2} + \mu_{2}^{2}k_{2}^{2}\right)\Gamma_{8} 
 +  {\rm i}\bigg[\left(\mu_{1}k_{1}^{3} + \mu_{2}k_{2}^{3}\right)\Gamma_{9}  
  + k_{1}k_{2}\left(\mu_{1}k_{2} + \mu_{2}k_{1}\right)\Gamma_{11}   
\nonumber \\&&{}
~~~~+ \left(\mu_{1}^{3}k_{1}^{3}+ \mu_{2}^{3}k_{2}^{3}\right)\Gamma_{12}
 + \mu_{1}\mu_{2}k_{1}k_{2}\left(\mu_{1}k_{1} + \mu_{2}k_{2}\right)\Gamma_{13} \bigg] \bigg\}\,,
\end{eqnarray}
where the $\Gamma_I(z)$ are given in Appendix \ref{sec:appendixB}.

Using the above, we find that the observed galaxy bispectrum in the squeezed limit is given by 
\begin{eqnarray}
{B_{g}(k_L,k_S, \mu_S,  {\phi})\over 2P(k_S)P(k_L)}
&=& b_{1S}b_{1L}b_{SL}+ \bigg[ b_{1S}\big(b_{SL}\gamma_2-f\gamma_1^2\mu_S^2\mu_L^2\big) +b_{1S}b_{1L}\big(\Gamma_7+\Gamma_{8}\mu_S^2\big)-b_{1L}\gamma_1\big(\Gamma_9+\Gamma_{12}\mu_S^2\big)\mu_S^2
 \nonumber\\
&&~~~~~~~~~~~~~~ -b_{1S}\gamma_1\big(\Gamma_{11}+\Gamma_{13}\mu_S^2\big)\mu_L^2\bigg]{1\over k_L^2}
+\gamma_2\bigg[b_{1S}\big(\Gamma_7+\Gamma_{8}\mu_S^2\big)-\gamma_1 \big(\Gamma_9+\Gamma_{12}\mu_S^2\big)\mu_S^2\bigg]{1\over k_L^4},
 \label{eq:bgs}
\end{eqnarray}
where we made the following definitions to simplify the expression:
\begin{equation}
b_{1S,L}\equiv b_1+f\mu_{S,L}^2,\qquad b_{SL}\equiv {10\over7}b_1+b_2
+ {6\over7}f\mu_S^2+b_1f\big(\mu_S^2+\mu_L^2\big)+2{f^2}\mu_S^2\mu_L^2 \,.
\end{equation}
For simplicity, we focus on the monopole contribution to $B_{g}$,
\begin{eqnarray}
B_{g}^{\,0}(k_L,k_S)&=& \frac{1}{4\pi}\int_0^{2\pi} d\phi\int_{-1}^{1}d\mu_{S}\, B_g(k_L,k_S, \mu_S,  {\phi})\nonumber\\
& =& \bigg({\cal B}_0+{{\cal B}_2\over k_L^2} + {{\cal B}_4\over k_L^4} \bigg)P(k_S)P(k_L),
\label{eq:bgsm}
\end{eqnarray}
where
\begin{eqnarray}
{\cal B}_0 &=&{ {20\over7}b_1^3+{52\over21}b_1^2f+2b_1^2b_2+{4\over3}b_1b_2f+ {68\over105}b_1f^2 + {4\over3}b_1^3f+{4\over3}b_1^2f^2 + {12\over35}b_1f^3+{2\over15}b_2f^2+{12\over 245}f^3+{4\over105}f^4 }\,, \label{ca0}\\
{\cal B}_2 &=& 
 {2\over105}\Bigg\{{\gamma_2}\Big(80b_1f+42b_1f^2+
 150b_1^2 +70b_1^2f  +105b_1b_2+ 35b_2f+18f^2 + 6f^3\Big)
+7b_1\Big[5\big(3b_1+f \big)
\Gamma_7+ \big(5b_1+3f \big)\Gamma_{8}\Big] \nonumber\\
&&-\gamma_1\Big[\big(35b_1+7f\big)\Gamma_{11}+\big(7b_1+3f\big)\Gamma_{13}\Big]+f\Big[\big(35b_1+7f\big)\Gamma_{7}+\big(7b_1+3f\big)\Gamma_{8}\Big]
\nonumber\\
    &&{}- {\gamma_1}\Big[\big(7b_1+3f \big)f\gamma_1+7\big(5b_1+f\big)\Gamma_9+3\big(7b_1+f\big)\Gamma_{12}\Big]   \Bigg\} \,,
\label{ca2} \\
{\cal B}_4 &=& {2\over15} \gamma_2  \Big[5\big(3b_1+f\big)\Gamma_7+\big(5b_1+3f\big)\Gamma_{8}
-{\gamma_1}\big(5\Gamma_9 + 3 \Gamma_{12}\big)
\Big] \,.
  \label{ca4} 
\end{eqnarray}
The Newtonian contribution to the squeezed bispectrum monopole is ${\cal B}_0$. 

Now we are in a position to compare the monopole of the relativistic galaxy bispectrum in a universe  with  Gaussian initial conditions, \eqref{eq:bgsm}, to the monopole of the Newtonian galaxy bispectrum with primordial local non-Gaussian initial conditions. (A similar comparison for the power spectrum was made in \cite{Jeong:2011as}.)
The local non-Gaussianity is defined in terms of the primordial gravitational potential by
\begin{equation}
\Phi =  {\varphi_{\rm G}} + \fnl \Big[ \varphi_{\rm G}^2
- \big\langle \varphi_{\rm G}^2 \rangle \Big],
\end{equation}
where $\varphi_{\rm G}$ is the Gaussian gravitational potential, related to the linear density contrast  {in CS gauge} by 
\begin{eqnarray}
\varphi_{\rm G}( {\k}) = \frac{\delta_{\rm G}( {\k},z)}{\alpha(k,z)}\,, \qquad \alpha(k,z) \equiv  {-}\frac{2 k^2 T(k) D(z)}{3 \Omega_{m0} H_0^2} = {-}\frac{2 g(z) T(k)}{3 \Omega_m( {z}) \Ha^2( {z})} k^2 \,.
\end{eqnarray}

 In the case of the power spectrum at linear order, $\fnl$ arises only in the scale-dependent linear galaxy bias, while for the bispectrum, $\fnl$ arises in \cite{Baldauf:2010vn, Tellarini:2016sgp}  
\begin{itemize}
\item the second-order matter density contrast, via   {$\fnl \big[\delta^{(1)}(\k,z)\big]{}^2\alpha(k,z)/\big[\alpha(k_1,z)\alpha(k_2,z)\big]$};
\item the second-order peculiar velocity, via    {$\fnl f(z) \big[\delta^{(1)}(\k,z)\big]{}^2\alpha(k,z)/\big[\alpha(k_1,z)\alpha(k_2,z)\big]$};
\item   {the scale-dependent galaxy bias:} 
\begin{eqnarray}
\delta_g =  b_1 \delta  + b_{01} \varphi_{\rm G} 
+ b_2\,  {\delta^2} + b_{11}\,  {\varphi_{\rm G}\delta} + b_{02}\, {\varphi_{\rm G}^2 } - b_{01}  N^2 \,,
\end{eqnarray}
where the non-Gaussian shift $N^2$ describes the effect of the change in gravitational potential due to a shift in galaxy positions from the initial position in the Lagrangian frame \cite{Tellarini:2016sgp}.
The new bias parameters are $b_{01} = \fnl c_{01}$, $ b_{11} = \fnl c_{11}$ and $b_{02} = \fnl^2 c_{02}$, where $c_{ij}$ depend on $\delta_c (=1.68)$, $b_1$ and $b_2$:
\begin{eqnarray}
c_{01}&=&2 \delta_c \left(b_{1}-1\right)\,,\\
c_{11}&=&2 \left[\delta_c b_{2}+\left(\frac{13}{21}\delta_c-1\right)\left(b_{1}-1\right)\right]\,,\\
c_{02}&=&4\delta_c\left[\delta_c b_{2}-2\left(\frac{4}{21}\delta_c+1\right)\left(b_{1}-1\right)\right]\,.
\end{eqnarray}
\end{itemize}

\begin{figure}[!ht]
\centering
\includegraphics[width=0.75\textwidth] {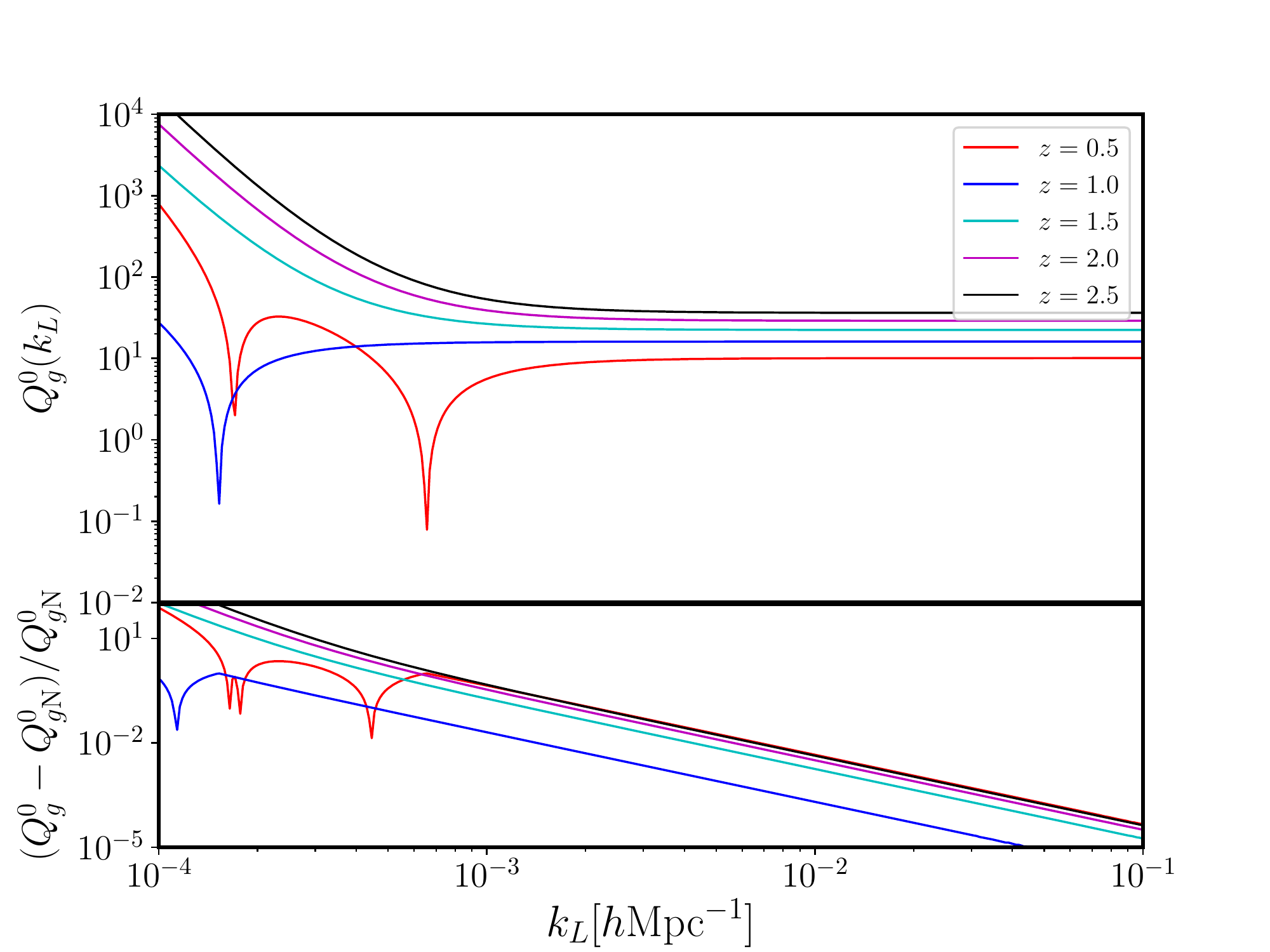}
\caption{Reduced monopole of the relativistic galaxy bispectrum at various redshifts ({\em top}), and the difference relative to the Newtonian approximation (Gaussian case) ({\em bottom}).  
}
\label{fig:f1}
\end{figure}
\begin{figure}[! h]
\centering
\includegraphics[width=0.75\textwidth] {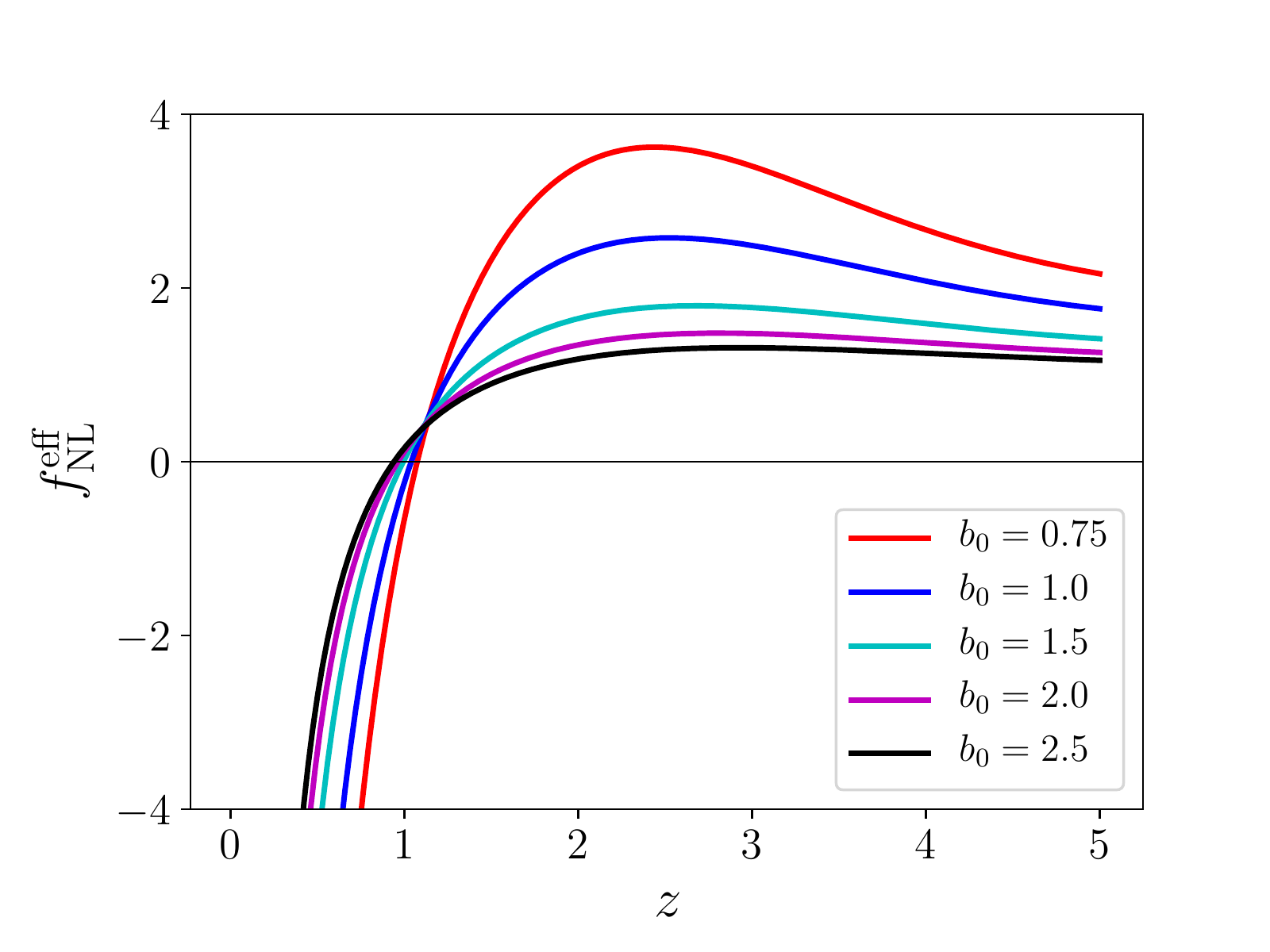}
\caption{Redshift dependence of the effective non-Gaussianity due to relativistic light-cone effects, for various values of $b_0$. 
}
\label{fig:f2}
\end{figure}

The monopole of the  {Newtonian} galaxy bispectrum with $\fnl \neq 0$  in the squeezed limit is  \cite{Tellarini:2016sgp}
\begin{eqnarray}
B_{g  {\rm N}}^{0}(k_L,k_S)\Big|_{\fnl}&=&
  \bigg( {{\cal B}_{{\rm N}0}}+\fnl { {{\cal B}_{{\rm N}2}}\over k_L^2} + \fnl^2{ {{\cal B}_{{\rm N}4}}\over k_L^4} \bigg)P(k_S)P(k_L),
\label{eq:Newtonianbispectrumsq}
\end{eqnarray}
where $ {{\cal B}_{{\rm N}0}} = {{\cal B}_0}$ and
\begin{eqnarray}
 {{\cal B}_{{\rm N}2}}&=&\frac{2 }{105}b_1 \bigg[\left(6 \beta^3 + 54 \beta^2 + 210 \beta + 210 \right)b_1^2 + 3 \beta\left(3 \beta^2 + 28 \beta + 35  \right) b_1^2 c_{01}
\\ \nonumber &&
+ 7 \left(\beta^2 + 10 \beta + 15 \right) b_1 c_{11} + 70\left(\beta + 3 \right) b_2 c_{01} \bigg]\frac{3 \Omega_{m} \Ha^2}{2 g}  \,, \qquad \beta \equiv \frac{f}{b_1}\,,
\\
 {{\cal B}_{{\rm N}4}}&=&\frac{2 }{105}b_1c_{01}\bigg[\beta\left(3 \beta + 5 \right) b_1 \left(28 + c_{01}\right) 
+ 35 \left(\beta + 3 \right) c_{11} \bigg]\left(\frac{3 \Omega_{m} \Ha^2}{2 g}\right)^2\,.
\end{eqnarray}

It is sufficient to compare  \eqref{eq:Newtonianbispectrumsq} and \eqref{eq:bgsm} at order $1/ k_L^2$, since the $1/ k_L^4$ contribution is important only on super-horizon scales.  {Then the effective non-Gaussianity contributed by local relativistic projection effects is given by}
\begin{eqnarray}
\fnl^{\rm{eff}} &=& {{\cal B}_2\over {\cal B}_{{\rm N}2}}.
\label{eq:efffnl}
\end{eqnarray}
In order to compute this effective parameter, we need to specify the astrophysical parameters $b_1,b_2,b_e, {\cal Q}$. At first order,  {the bias of many tracers can be modelled by} a simple square-root dependence on redshift \cite{Sheth:1999mn}. At second order,  using a halo model technique proposed in \cite{Tellarini:2016sgp}, we can relate the two bias parameters. This leads to 
\begin{eqnarray}
b_1(z)=b_0 \sqrt{1+z}\,,\qquad {b_2(z)=   -0.25\,b_0-0.13\,\sqrt{1+z}\,\sin(0.8 z) } \,,
\end{eqnarray}
where $b_0$ is the only free parameter. We choose $b_0=1$.
For simplicity, we set the magnification and evolution bias parameters to zero: ${\cal Q}=0=b_e$. 

Figure \ref{fig:f1} (top panel) shows the reduced relativistic galaxy bispectrum in the squeezed limit,
\begin{equation}
Q^0_g (k_L) = \frac{B^0_g (k_L, k_S) }{P(k_S) P(k_L)}\,,
\end{equation}
at various redshifts, where the fractional difference relative to the Newtonian $ {Q^0_{g \rm N}}$ is shown in the bottom  panel. Lightcone effects induce corrections to the Newtonian bispectrum at $> 1\,$\%  when the long mode  is $\lesssim 5\times 10^{-3} h\,{\rm Mpc}^{-1}$ (depending on $z$), and this grows for smaller $k_L$.  

The redshift dependence of $\fnl^{\rm{eff}}$, defined in  \eqref{eq:efffnl}, is shown in Fig. \ref{fig:f2}, for varying bias parameter $b_0$. 
It is very sensitive to the galaxy bias, and also depends on the evolution and magnification bias. The shape of the $\fnl^{\rm{eff}}$ curve is similar to that given in  \cite{Kehagias:2015tda} for their $t=-2$ (our ${\cal Q}=0)$ case, but note that they neglect galaxy bias and its evolution by assuming $b_1=1=b_2$.
We should emphasise that in order to obtain an accurate estimate of $\fnl^{\rm{eff}}$, we need to project the full bispectrum onto the bispectrum arising from primordial local non-Gaussianity. This was done by \cite{DiDio:2016gpd}, using a subset of the contributions to the bispectrum as well as the consistency relation obtained in \cite{Kehagias:2013yd}. They found $\fnl^{\rm{eff}} = {\cal O}(1)$, which is consistent with the high redshift behaviour in Fig. \ref{fig:f2} -- noting that  \cite{DiDio:2016gpd} also neglects galaxy bias.

\section{Conclusion}\label{sec:conc}
We derived the consistency relation for the second-order observed galaxy number density contrast. The effect of long-mode perturbations on small-scale modes is to modulate the observed coordinates of small-scale perturbations. They also introduce perturbations in observed luminosity and volume. Using this simple prescription, we were able to derive the second-order observed galaxy number density fluctuations as a product of long- and short-mode perturbations. This gives the bispectrum in the squeezed limit and it is known as the consistency relation. 

The second-order observed galaxy density contrast has been computed using second-order cosmological perturbations in the literature. There is a subtlety in the computation which we highlighted and resolved. The long-mode curvature perturbation introduces a coupling between long and short modes in the second-order dark matter density and velocity. This is the consistency relation in real space. However, this is purely a gauge effect  {in a single-field inflation universe}, since the long-mode curvature perturbation can be absorbed into the background coordinates in a small patch. The effect of the long mode reappears if we consider correlation between different patches. This is what is captured in the consistency relation for the observed galaxy perturbations. In the second-order computation, this is included in the projection effect. Thus we need to remove the squeezed limit contribution to the second-order dark matter perturbations that is generated by the spatial dilatation caused by long-mode curvature perturbations. We confirmed this by explicitly comparing the second-order result in the squeezed limit and the consistency relation. 

The consistency relation provides a useful way to confirm the full second-order computation. It also serves as a consistency check for second-order numerical codes, as was shown for the cosmic microwave background bispectrum \cite{Creminelli:2004pv,Creminelli:2011sq,Bartolo:2011wb,Lewis:2012tc,Pettinari:2014iha}. 

Finally, we quantified the effective local non-Gaussianity due to the relativistic lightcone projection effects in the squeezed limit, which could potentially contaminate the primordial non-Gaussianity signal if not properly accounted for.

\[\]
{\bf Acknowledgments:}\\
{We thank Cornelius Rampf and Eleonora Villa for clarification on the second-order peculiar velocity in Poisson gauge. Some of the algebraic computations here were done with the tensor algebra software xPand \cite{Pitrou:2013hga}}.
KK, OU and RM are supported by the UK STFC grant ST/N000668/1.  KK is also supported by the European Research Council under the European Union's Horizon 2020 programme (grant agreement 646702 ``CosTesGrav"). RM is also supported by the South African SKA Project and the National Research Foundation (Grant No. 75415). DB acknowledges partial financial support by ASI Grant No. 2016-24-H.0. During the preparation of this work DB was also supported by the Deutsche Forschungsgemeinschaft through the Transregio 33, The Dark Universe and Unidad de Excelencia ``Mar\'ia de Maeztu".

\newpage
\begin{appendix}
\section{Integral terms}\label{sec:appendixA}
Here we derive $\Delta x_{\|}$ in the CS gauge in an alternative way and discuss its gauge dependence. The line element is 
\begin{equation}
ds^2 = a^2(\eta) \Big\{ - d \eta^2 + \Big[ (1+2 D) \delta_{ij} + 2 E_{ij}\Big] dx^i dx^j \Big\},
\end{equation}
 {where $E_{ij} =\big(\partial_i \partial_j - \frac{1}{3} \delta_{ij} \nabla^2 \big) E.$}
In the CS gauge, 
\begin{equation}
\Delta x_{\|}^{\rm CS} = -\frac{1}{\Ha} \delta z^{\rm CS} - T^{\rm CS}, 
\label{deltaxCS}
\end{equation}
where 
\begin{equation}
T^{\rm CS} = \int^{\chi}_0  d \chi' (D + E_{\|}), \quad \delta z^{\rm CS} = \int^{\chi}_0 d \chi'  \Big(D' + E_{\|}' \Big), \quad 
E_{\|} = \partial^2_{\|} E - \frac{1}{3} \nabla^2 E \,.
\end{equation}
We express these in terms of metric perturbations in the Poisson gauge
\begin{equation}
ds^2 = a^2(\eta) \Big[- (1 + 2 \Psi) d \eta^2 +(1 - 2 \Phi) \delta_{ij} dx^i dx^j \Big]\,,
\end{equation}
where
\begin{equation}
\Psi =  - \Ha E' - E'', \quad \Phi = - D + \frac{1}{3} \nabla^2 E + \Ha E', \quad v=E' \,.
\end{equation}
Then \eqref{deltaxCS} can be written  as 
\begin{equation}
\Delta x_{\|}^{\rm CS} = -\frac{1}{\Ha} \delta z^{\rm P} - T^{\rm P}  - \partial_{\|} \xi \,,
\label{deltaxCS2}
\end{equation}
where 
\begin{equation}
\delta z^{\rm P} = -\Phi + \partial_{\|} v + 2 I , 
\quad I^{\rm P} = - \frac{1}{2} \int_0^{\chi} d \chi' (\Phi' + \Psi'),
\quad T^{\rm P} = -  \int_0^\chi d \chi' (\Phi + \Psi), \quad \xi = \int d \eta' \;  v(\eta').  \\
\end{equation}
Here we used the fact that the time-delay and redshift perturbations in the CS and Poisson gauges are related as
\begin{eqnarray}  
	T^{\rm CS} &=& T^{\rm P}  - v + \partial_{\|} \xi \,,	\label{timedelay}\\
\delta z^{\rm CS} &=& \delta z^{\rm P} - \Ha v \,.
\end{eqnarray}
Then we obtain 
\begin{align} 
	\Delta x_{\|}^{\rm CS} =\Delta x_{\|}^{\rm P} - \partial_{\|} \xi, \quad 	 
	\Delta x_{\|}^{\rm P} = -\frac{1}{\Ha} \delta z^{\rm P} - T^{\rm P}. 
\end{align}
This is an expected result from the gauge transformation  {$x^{{\rm CS}i} 
=x^{ {\rm P}i} - \xi^{,i}$}, where $\xi$ is the spatial gauge transformation.

In the main text, we ignore the time delay term $T^{\rm P}$ in (\ref{deltaxCS2}) as an integral term. This is not equivalent to ignoring the time delay term $T^{\rm CS}$ in (\ref{deltaxCS}), since the time-delay term in the CS gauge contains a local velocity term, shown in (\ref{timedelay}). In \cite{Kehagias:2015tda}, it appears that the time-delay term in the CS gauge was ignored instead of that in the Poisson gauge in (\ref{deltaxCS}). This is the reason that our expressions for $\Delta x_{\|}$ (and $ {\Delta x_{\perp}^i}$) are different from those given in \cite{Kehagias:2015tda}.

\section{Fourier kernels}\label{sec:appendixB}

For completeness, we present the redshift-dependent terms in the squeezed limit bispectrum kernels. The complete list of these terms can be found in \cite{Jolicoeur:2017nyt}. At first order, the coefficients in \eqref{ker1} are
\begin{eqnarray}
{\gamma_1\over \mathcal{H}} &=&  { f}\bigg[b_{e}  - 2\mathcal{Q} - \frac{2(1-\mathcal{Q})}{\chi\mathcal{H}}- \frac{\mathcal{H}'}{\mathcal{H}^{2}} \bigg] , \label{ga1}
\\
\frac{\gamma_{2}}{\mathcal{H}^{2}} &=& f(3-b_{e}) +\frac{3}{2}\Omega_{m} \left[2+b_{e}-f-4\mathcal{Q}-2 \frac{\left(1- \mathcal{Q} \right)}{\chi\mathcal{H}}-\frac{\mathcal{H}'}{\mathcal{H}^2}\right]. \label{ga2}
\end{eqnarray}
At second order, the Fourier expansion of Newtonian approximation is (suppressing redshift dependence):
\begin{align}
 {\Delta_{g\rm N}^{(2)}({\k})} &= \int \frac{d^{3}k_{1}}{(2\pi)^{3}}\int d^{3}k_{2}\,
{\mathcal{K}^{(2)}_{\rm N}( {\k}_{1}, {\k}_{2}, \k)}\delta^{(1)}( {\k}_{1})\delta^{(1)}( {\k}_{2})\delta^{D}({\k}_{1} + {\k}_{2} - {\k})\,.
\end{align}
The kernel is given in \eqref{eq:FourierNewtonian}, where
\cite{Bernardeau:2001qr}
\begin{eqnarray} 
F_{2}(\bm{k}_{1}, \bm{k}_{2}) &=& \frac{10}{7} + \frac{\bm{k}_{1} \cdot \bm{k}_{2}}{k_{1}k_{2}}\bigg(\frac{k_{1}}{k_{2}} + \frac{k_{2}}{k_{1}}\bigg) + \frac{4}{7}\bigg(\frac{\bm{k}_{1} \cdot \bm{k}_{2}}{k_{1}k_{2}}\bigg)^{2}  \,, \\
G_{2}(\bm{k}_{1}, \bm{k}_{2}) &=& \frac{6}{7} + \frac{\bm{k}_{1} \cdot \bm{k}_{2}}{k_{1}k_{2}}\bigg(\frac{k_{1}}{k_{2}} + \frac{k_{2}}{k_{1}}\bigg) + \frac{8}{7}\bigg(\frac{\bm{k}_{1} \cdot \bm{k}_{2}}{k_{1}k_{2}}\bigg)^{2}.
\end{eqnarray}
 The $\Gamma_I$ in \eqref{eq:kSL} are 
\begin{eqnarray}
\frac{\Gamma_{7}}{\Ha^{2}} &=& \frac{3}{2}\Omega_{m}\Bigg[b_{1}\bigg(2+b_{e}-4\Q-\frac{2(1-\Q)}{\chi\Ha} -\frac{\Ha'}{\Ha^{2}}\bigg) + \frac{b_{1}'}{\Ha} + 2\bigg(2-\frac{1}{\chi\Ha}\bigg)\frac{\p b_{1}}{\p \ln{\bar{L}}}\Bigg] - f\Bigg[b_{1}(f-3+b_{e}) + \frac{b_{1}'}{\Ha}\Bigg],  \\
\frac{\Gamma_{8}}{\Ha^{2}} &=& \frac{9}{4}\Omega_{m}^{2} + \frac{3}{2}\Omega_{m}f\Bigg[{1}-2f+2b_{e}-{6}\Q-\frac{4(1-\Q)}{\chi\Ha}-\frac{3\Ha'}{\Ha^{2}}\Bigg] + f^{2}(3-b_{e}) +2\left(f^2-\frac{3}{4} \Omega_m f\right),\\ \nonumber \\ \nonumber
\frac{\Gamma_{9}}{\Ha} &=& -\frac{3}{2}\Omega_{m}b_{1}, \\
\frac{\Gamma_{11}}{\Ha} &=& f\Bigg[b_{1}\bigg(f+b_{e}-2\Q-\frac{2(1-\Q)}{\chi\Ha}-\frac{\Ha'}{\Ha^{2}}\bigg) + \frac{b_{1}'}{\Ha} + 2\bigg(1-\frac{1}{\chi\Ha}\bigg)\frac{\p b_{1}}{\p \ln \bar{L}}\Bigg], \\
\frac{\Gamma_{12}}{\Ha} &=& -\frac{3}{2}\Omega_{m}f, \\
\frac{\Gamma_{13}}{\Ha} &=& \frac{3}{2}\Omega_{m}f -f^{2}\Bigg[3-2b_{e}+{4}\Q+\frac{4(1-\Q)}{\chi\Ha}+\frac{3\Ha'}{\Ha^{2}}\Bigg]. 
\end{eqnarray}
Here $\Gamma_8$ includes the contribution from the intrinsic second-order peculiar velocity term with the gauge mode subtracted out, as described in section \ref{sec:spatialdilatation}.
\end{appendix}

\bibliography{Symmetry-bib-file}

\end{document}